\documentclass[aps,prb,preprint,floatfix,superscriptaddress,showpacs,amssymb,amsmath]{revtex4}

\usepackage{graphicx}

\begin{document}

\title{Magnetic semiconductor artificial atom with many particles: Thomas-Fermi
model and ferromagnetic phases}
\author{Alexander O. Govorov}
\affiliation{Department of Physics and Astronomy, Condensed Matter
and Surface Science Program, \\ Ohio University, Athens, Ohio
45701-2979}

\date{\today } %10/22/03

\begin{abstract}
Many-particle electron states in semiconductor quantum dots with
carrier-mediated ferromagnetism are studied theoretically within
the self-consistent Boltzmann equation formalism. Depending on the
conditions, a quantum dot may contain there phases: partially
spin-polarized ferromagnetic, fully spin-polarized ferromagnetic,
and paramagnetic phases. The physical properties of many-body
ferromagnetic confined systems come from the competing
carrier-mediated ferromagnetic and Coulomb interactions. The
magnetic phases in gated quantum dots with holes can be controlled
by the voltage or via optical methods.
\end{abstract}


%Optical properties of low-dimensional, mesoscopic, and nanoscale materials and structures
%78.67.Hc Quantum dots

% 73.21.-b    Electron states and collective excitations in
%multilayers, quantum wells, mesoscopic, and nanoscale systems
%(for electron states in nanoscale materials, see 73.22.-f)

%73.21.La    Quantum dots

%75.75.+a Magnetic properties of nanostructures

%73.63.-b    Electronic transport in nanoscale materials and structures
%73.63.Kv    Quantum dots

 \maketitle

\section{ Introduction }

Diluted magnetic semiconductors \cite{Furdyna} represent an
important class of materials and structures where ferromagnetism
can be tuned by voltage \cite{S}. This ability comes from the
carrier-mediated character of the ferromagnetic interaction
\cite{Zener,M,D}. The ferromagnetic ordered state in these systems
appears due to mobile carriers interacting with stationary spins
of magnetic impurities. To date, Curie temperatures as high as
$40~K$ have been observed in a technologically important class of
the Mn-doped III-V semiconductor structures
\cite{Giddings,Burch,Molenkamp}. When the magnetic semiconductors
become combined with the conventional field-effect layered
structures, the number of mobile carriers and the ferromagnetic
interaction become tunable by the voltage \cite{VControlFerro}.
This ability to externally control the properties of magnetic
crystals with means other than the external magnetic field may
have important device applications. A further step from magnetic
semiconductor layers would be zero dimensional systems, quantum
dots (QDs). Magnetic quantum dots can be viewed  as nano-scale
memory elements where information is stored in the form of
magnetic polarization. Such a system may have important advantages
compared to the conventional metal spin-valve memories: (1) small
sizes and relatively small number of carriers and (2) voltage
control of the number of electrons which was already demonstrated
in many experiments for non-magnetic QDs
\cite{QDVoltage1,QDVoltage2}. Therefore it is important to develop
the understanding of magnetic QDs with interacting carriers.

Here we develop a theory of magnetic QDs with many carriers where
Coulomb, ferromagnetic, and single-particle energies contribute to
the formation of the equilibrium state. Using the quasi-classical
description, we show that a QD may be split into three phases with
different physical properties. The geometrical sizes of these
phases are determined by the Coulomb, ferromagnetic, and
single-particle contributions to the chemical potential of a QD.
For calculations, we employ the mean-field theory and the
Boltzmann kinetic equation. This approach becomes reduced to the
Thomas-Fermi model at low temperatures. We should note that our
approach ignores the discrete structure of single-particle
spectrum of QD and is valid when electrons occupy at least several
quantum levels. At the same time, this approach has an important
advantage: it allows us to describe the Coulomb effects in
relatively large QDs where, as it is shown below, the Coulomb
interaction is very strong and significantly exceeds the
ferromagnetic interaction and the kinetic energy of carriers. The
hole-mediated ferromagnetism in quasi-two-dimensional (2D) systems
is strongly anisotropic due to the heavy hole-light hole splitting
in the valence band. Therefore, the magnetic polarization occurs
predominantly in the growth direction. Then, two magnetic states
of a QD with spin polarizations "up" and "down" can represent a
single bit.

Presently QDs and other nano-structures doped with magnetic (Mn)
impurities attract a lot of attention. Among other studies,
several recent theoretical papers investigate QDs and their
electron and excitonic states in the presence of a single Mn ion
\cite{France,GovorovPRB04Mn,GovorovPRB05Mn2D,Pawel}. In
particular, it was suggested in ref.~\cite{GovorovPRB05Mn2D} that
a QD with a single Mn ion can act as a multi-qubit which can be
controlled optically. Another direction of research describes the
magnetic states and polarons in QDs with many Mn ions and one or
several carriers
\cite{Bhattacharjee0,Brey,Peeters,Kulakovskii-PRB,Cincinnati,Govorov-PRB2005}.
Among the above publications, the paper \cite{Govorov-PRB2005}
demonstrates that the Coulomb-interaction effects in few-electron
QDs can determine a collective magnetic state of holes and Mn
spins. Ferromagnetism and spin separation in digital layered
structures and quantum wells were also studied in several
experimental \cite{Burch,Bruce} and theoretical publications
\cite{D,LSham,Dietl2D}.

\section{ Model }

\begin{figure}[tbp]
\includegraphics*[width=0.4\linewidth,angle=90]{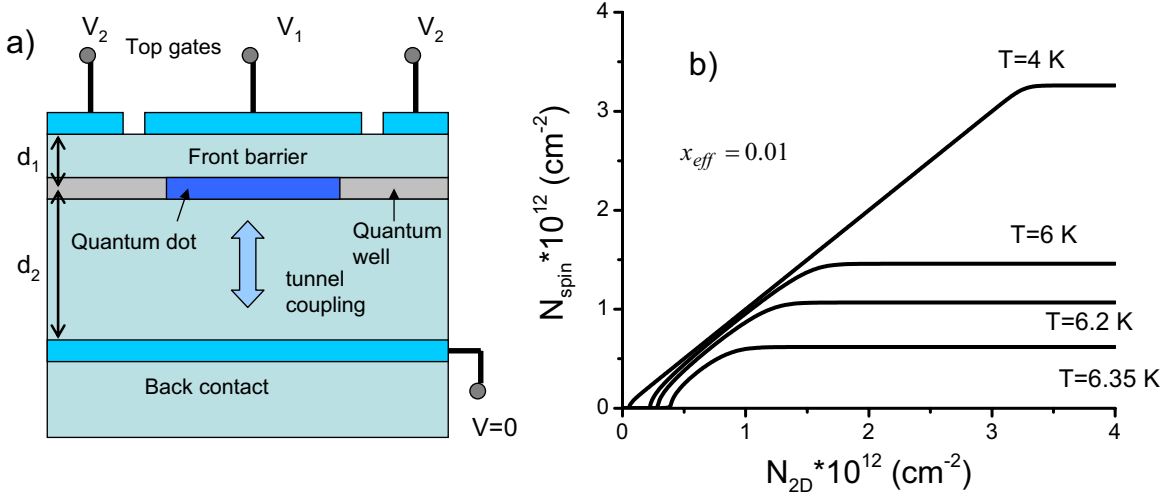}
\caption{(a) Model of a Mn-doped lithographically defined QD based
on a semiconductor quantum well. The number of particles is
controlled be the voltage applied between the top and back gates.
The QD confinement for holes is given by a voltage difference
$\Delta V=V_1-V_2<0$. (b) Calculated 2D spin density as function
of the total 2D density of holes at various temperatures;
$m_{hh}=0.38~m_0$, $L=70$ \AA, $N_0=23~nm^{-3}$, $\beta
N_0=-2.2~eV$, $x_{eff}=0.01$. Curie temperature for the above
parameters is about $6.4~K$.} \label{fig1}
\end{figure}

 As a model system, we are going to use a QD "made out
of" a 2D quantum well. Such a zero-dimensional system can be
fabricated by etching and lithographical methods. Within the
lithographical method,  a QD can be defined, for example, by using
the top metal gates (fig.~\ref{fig1}a). The number of carriers in
such a QD is a voltage tunable parameter.

To describe a state of many carriers confined in a QD, we start
from the local properties of the coupled hole-Mn system in a 2D
quantum well. In our system, a mobile hole and Mn spins experience
an exchange interaction: ${\hat
U}_{exc}=-\beta/3\sum_i(\hat{S}_{z,i}\hat{j}_z)\delta({\bf
r}_h-{\bf R}_{i})$, where $\beta$ is the exchange interaction
constant, and $\hat{S}_z$ and $\hat{j}_z$ are the z-components of
the spin operators for a Mn spin and hole, respectively; ${\bf
r}_h$ and ${\bf R}_{i}$ are the coordinates of hole and
$i$-impurity, respectively. The above operator describes the
interaction between Mn-spins and heavy holes and assumes a
sufficiently large energy separation between the heavy-hole and
light-hole quantized states in the valence band. The corresponding
effective spin-dependent potential of a single hole has a form:

\begin{eqnarray}
U_{spin}(j_z)=\frac{j_z}{|j_z|}U_{spin}^0 \label{Uspin},
\end{eqnarray}
where
\begin{eqnarray} \label{Uspin0}
U_{spin}^0(N_{spin})=-\frac{\beta}{3}x_{eff} N_0
\int{dz\psi_0^2(z)S_{Mn}B_{S_{Mn}}[\frac{-\beta}{3}\frac{3}{2}N_{spin}\psi_0^2(z)]}.
\end{eqnarray}
Here $N_{spin}=N_{+3/2}-N_{-3/2}$ is the net spin 2D density,
$N_{j_z}$ are the 2D densities of the spin components,
$B_{S_{Mn}}$ is the Brillouin function, $N_0$ is the number of
cations per unit cell, and $S_{Mn}=5/2$.
$N_{2D}=N_{+3/2}+N_{-3/2}$ is the total 2D density in a quantum
well. For the ground-state wave function in a quantum well, we
will use $\psi_0(z)=\sqrt{2/L}\sin{\pi z/L}$, where $L$ is the
quantum well width. The chemical potential for a 2D gas depends on
$U_{spin}^0$ and $N_{2D}$:

\begin{eqnarray}
\mu_{2D}(T,N_{2D},U_{spin})=k_BT
ln[-cosh(\frac{U_{spin}^0}{k_BT})+\sqrt{cosh(\frac{U_{spin}^0}{k_BT})^2+exp(\frac{2E_f}{k_BT})-1}],
\end{eqnarray}
where $E_f(N_{2D})=\pi N_{2D}\hbar^2/m_{hh}$. Now we calculate the
spin polarization:

\begin{eqnarray}
N_{spin}=\frac{k_BTD_{2D}}{2}
ln[\frac{1+e^{\frac{\mu_{2D}-U_{spin}^0(N_{spin})}{k_BT}}}{1+e^{\frac{\mu_{2D}+U_{spin}^0(N_{spin})}{k_BT}}}],
\label{Nspin}
\end{eqnarray}
where $D_{2D}=m_{hh}/(\pi\hbar^2)$. The Zener ferromagnetic phase
transition occurs when eq.~\ref{Nspin} has a nonzero solution.
Fig.~\ref{fig1}b shows the data for spin density $N_{spin}$ for a
GaAs/AlGaAs quantum well with the following parameters:
$m_{hh}=0.38~m_0$, $L=70~$ \AA, $N_0=23~nm^{-3}$, $\beta
N_0=-2.2~eV$, $x_{eff}=0.01$. The above exchange parameter $\beta$
is comparable to that used in other publications on magnetic
semiconductors (see e.g. \cite{LSham}). Since the exchange
interaction is antiferromagnetic ($\beta<0$), the case
$N_{spin}>0$ corresponds to the negative average polarization of
Mn ions, $B_{S_{Mn}}<0$. Curie temperature can be analytically
calculated in the high-density limit: $k_B T_C=S(S+1)\beta^2
x_{eff} N_0 m_{hh}/(8\pi\hbar^2)$ ($k_BT_c\ll N_{tot}/D_{2D}$).

The in-plane potential in a lateral QD near its center can be
approximated by the parabolic function:

\begin{eqnarray}
e\phi_0({\bf r}) =U_0+\frac{m_{hh}\omega^2}{2}r^2, \label{parab}
\end{eqnarray}
where $e>0$ is the electron charge and $\omega$ is a
characteristic frequency of a confining potential. The potential
$U_0$ determines a depth of a lateral potential well. In a QD
defined by metal gates (fig.~\ref{fig1}a), the parameters $U_0$
and $\omega$ are functions of the gate voltages. In equilibrium,
the carrier distribution function, which satisfies the Boltzmann
equation, has a form:

\begin{eqnarray}
f(p,{\bf r},j_z)=\frac{1}{e^{\frac{p^2/2m_{hh}+e\phi({\bf
r})+U_{spin}(j_z,{\bf r})-\mu}{k_BT}}+1}, \label{ff}
\end{eqnarray}
where ${\bf r}=(x,y)$ is the lateral position vector and ${\bf p}$
is the in-plane momentum. The self-consistent scalar potential of
a hole is composed of two terms:

\begin{eqnarray}
e\phi({\bf r})=e\phi_0({\bf r})+U_{Coul}({\bf r}), \label{pot}
\end{eqnarray}
where $U_{Coul}({\bf r})$ is the electrostatic potential induced
by a non-uniform spatial distribution of carriers, $n_{2D}({\bf
r})=n_{+3/2}({\bf r})+n_{-3/2}({\bf r})$. In addition, the
distribution function (\ref{ff}) depends on the spin of hole
through the exchange interaction which is a function of the local
spin density, $n_{spin}({\bf r})=n_{+3/2}-n_{-3/2}$ (see
eqs.~\ref{Uspin},\ref{Uspin0}). At the same time, the function
$n_{spin}({\bf r})$ itself is determined by the total local
density of holes, $n_{2D}({\bf r})$, and is given by the numerical
solution of eq.~\ref{Nspin} (see the data in fig.~\ref{fig1}b).
Therefore, it is convenient to regard $U_{spin}$ as a function of
$n_{2D}$, i.e. $U_{spin}[j_z,{\bf
r}]=U_{spin}[j_z,n_{spin}(n_{2D}))]=U_{spin}[j_z,n_{2D}({\bf
r})]$. By integrating the function (\ref{ff}) over momenta we come
to two non-local non-linear equations for the densities
$n_{\pm3/2}({\bf r})$. Then, these equations can be solved for the
chemical potential and rewritten in the form resembling the
central equation of the Thomas-Fermi model:

\begin{eqnarray}
\mu=e\phi_0({\bf r})+U_{Coul}({\bf
r})+(j_z/|j_z|)U_{spin}^0[n_{tot}({\bf r})]+k_B T
\ln[e^{\frac{2\pi\hbar^2 n_{j_z}}{m_{hh} k_B T}}-1], \hskip 0.5cm
j_z=\pm3/2, \label{TF}
\end{eqnarray}
where $j_z=\pm3/2$ and $U_{Coul}({\bf r})=e^2\int d^2{\bf
r'}\frac{n_{2D}({\bf r})}{\epsilon_{eff}(|{\bf r'}-{\bf r}|) |{\bf
r'}-{\bf r}|}$, where $\epsilon_{eff}({\bf r'}-{\bf r})|$ is an
effective non-local dielectric constant of a system with metal
gates \cite{GovorovJETPLett}. We should also note that
$U_{Coul}({\bf r})$ was written as a 2D integral and this it valid
if the lateral size of a QD is greater than the quantum-well width
$L$.

In the system with the top gates closely located to a quantum
well, the 2D integral in the equation for $U_{Coul}({\bf r})$ is
reduced to a linear function of $n_{2D}$ \cite{GovorovJETPLett}
and is given by a local flat-capacitor formula:

\begin{eqnarray}
\label{FC} U_{Coul}({\bf r})=\frac{4\pi e^2  d_1  n_{2D}({\bf
r})}{\epsilon_s},
\end{eqnarray}
where $d_1$ is the distance between the QD plane and the top gate;
the distance to the back metal contact is assumed to be larger,
i.e. $d_2\gg d_1$; $\epsilon_s$ is a dielectric constant of the
semiconductor ($\epsilon_s=12.5$). The approximation (\ref{FC})
has been successfully used in the past for description of several
experiments on optical and electronic properties of modulated
lateral structures \cite{GovorovLateralExcitons,GovorovCapExper}.
By using the local approximation for the self-consistent potential
(\ref{FC}), we reduce eqs.~\ref{TF} to coupled non-linear local
equations which should be solved numerically. The total number of
holes in a QD is determined by the chemical potential $\mu$ and
the lateral-potential depth $U_0$. Electrostatics of the structure
under study (fig.~\ref{fig1}a) is similar to that studied in
refs.~\cite{GovorovLateralExcitons,GovorovCapExper,GovorovCapTheory}
and we can use here the results of the above publications. If the
barrier between the QD and back contact permits efficient
tunnelling, the chemical potential in the QD coincides with the
potential of the back contact (i.e. $\mu=0$). Simultaneously, the
front barrier (usually made of AlGaAs) blocks tunnelling between
the QD and the top gate. Also, if $d_1\ll d_2$, the potential
$U_0$ becomes close to $eV_1$.

\section{ Magnetic phases in quantum dots }

\begin{figure}[tbp]
\includegraphics*[width=0.6\linewidth,angle=90]{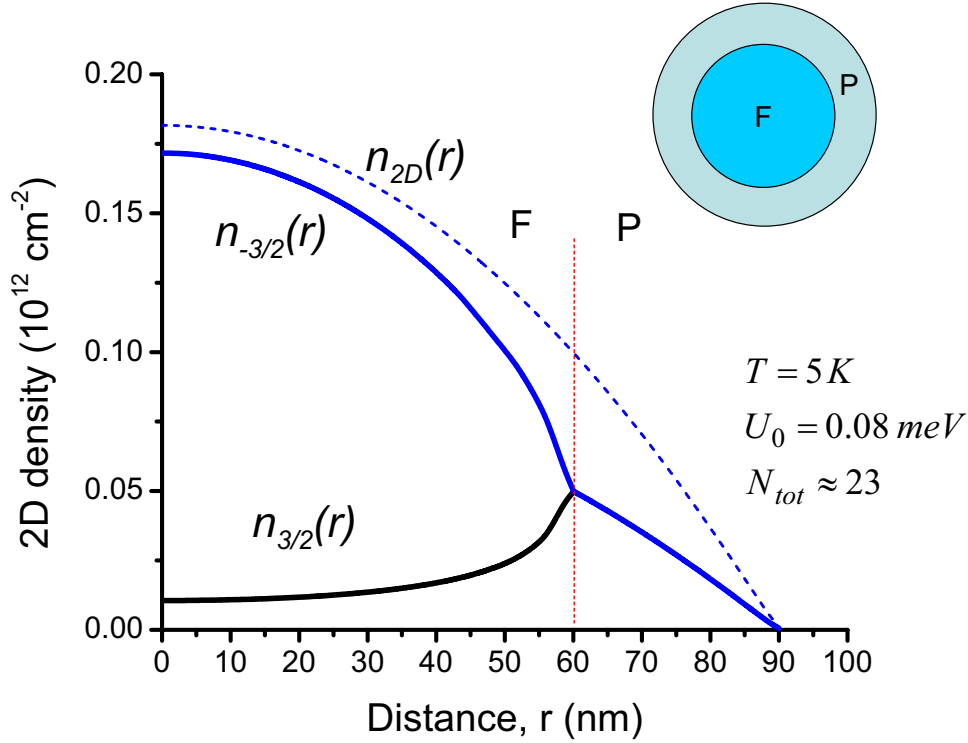}
\caption{Calculated hole density as a function of the distance
from the center of QD for a relatively small number of holes. The
QD is divided into ferromagnetic (F) and paramagnetic phases (P).
The dashed line represents the total 2D density.  Inset: sketch of
the QD structure.} \label{fig2}
\end{figure}

\begin{figure}[tbp]
\includegraphics*[width=0.6\linewidth,angle=90]{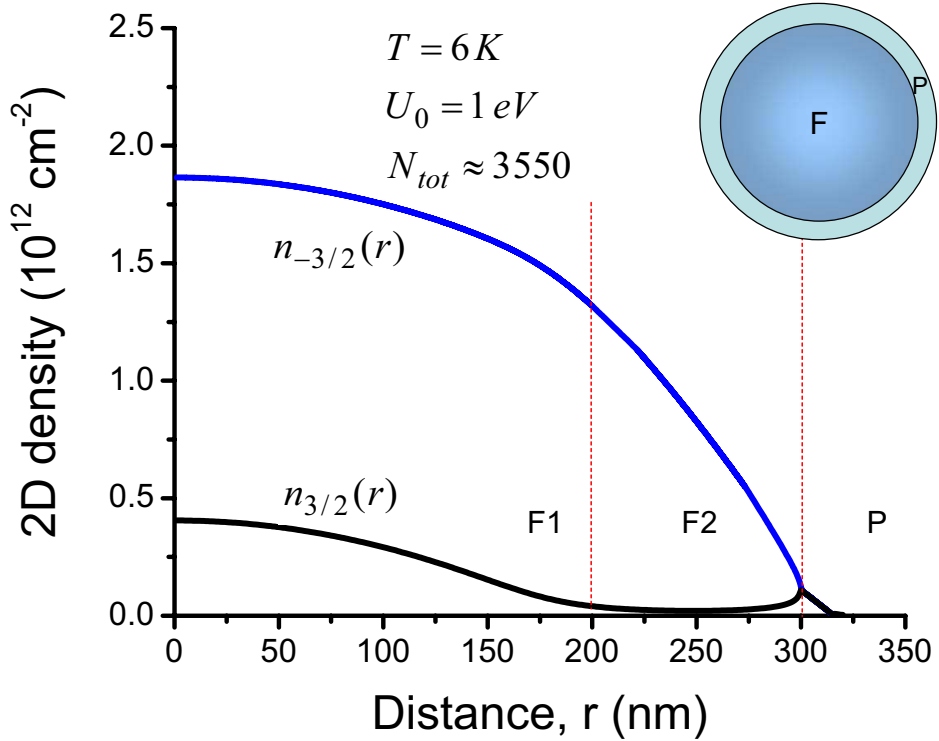}
\caption{(Calculated hole density as a function of the distance
from the center of QD for a larger number of holes. The QD is
divided into ferromagnetic (F1 and F2) and paramagnetic regions
(P). Inset: schematics of the QD structure.} \label{fig3}
\end{figure}

\begin{figure}[tbp]
\includegraphics*[width=0.6\linewidth,angle=90]{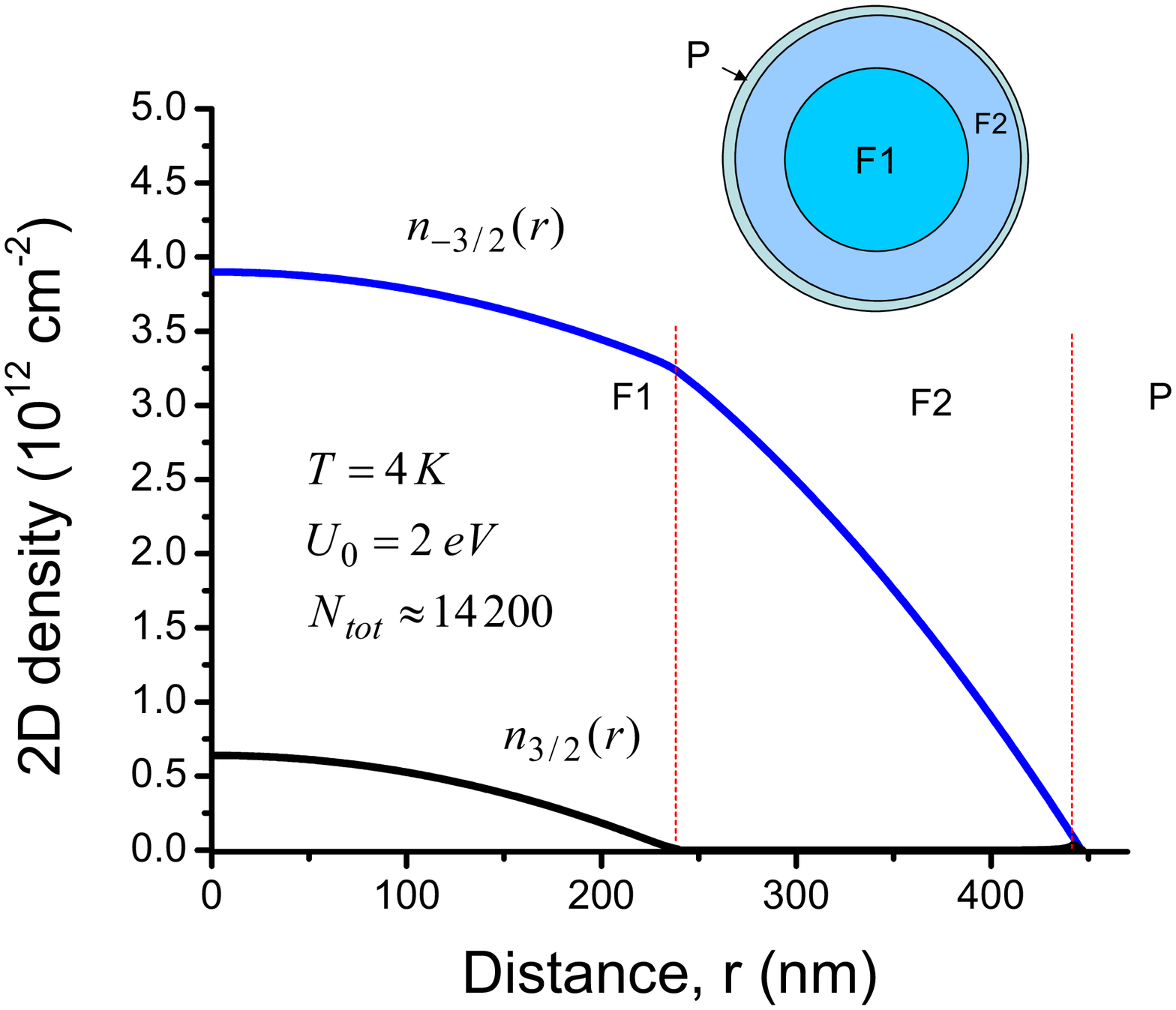}
\caption{Calculated hole density as a function of the distance
from the center of QD for a large number of holes at low
temperature. The QD is divided into ferromagnetic (F1 and F2) and
paramagnetic phases (P). Inset: schematics of the QD structure.}
\label{fig4}
\end{figure}

Figs.~\ref{fig2},\ref{fig3}, and \ref{fig4} show numerical
calculations for the local spin densities in a circular QD with
$d_1=300$ \AA, $\hbar\omega=2~meV$, and $\mu=0$; for the QD depth,
we take $U_0=-0.08, -1$, and $-2~eV$. A QD with the minimum free
energy is circularly symmetric and can be split into different
phases. The total number of holes in a QD is given by an integral
$N_{tot,QD}=\int n_{2D}(r) d^2{\bf r}$. The corresponding
$N_{tot,QD}$ for the above values of $U_0$ are estimated as $23,
3550$, and $14200$. In QDs with relatively small $N_{tot,QD}$
(fig.\ref{fig2}), the system is split into ferromagnetic (F) and
paramagnetic (P) phases. In fig.~\ref{fig2}, the carriers with
spin $j_z=+3/2$ are pushed away from the center of QD, the total
spin of holes is negative, and the Mn-subsystem has a positive
magnetization. This situation corresponds to the antiferromagnetic
hole-Mn coupling ($\beta<0$). With increasing the total number of
carriers (fig.~\ref{fig3}), one can see the formation of another
stripe within the ferromagnetic phase. This stripe is located
between the center region of a QD and the paramagnetic phase and
the holes in this stripe are almost fully spin-polarized. With
further increasing $N_{tot,QD}$ (fig.~\ref{fig4}) and for
relatively low temperatures, the formation of the ferromagnetic
stripe (F2) with filly-polarized holes becomes evident.
Simultaneously, the paramagnetic stripe becomes very narrow. Such
a magnetic stricture of a QD can be understood by looking at the
data in fig.~\ref{fig1}b. At low temperatures, the function
$N_{spin}(N_{2D})$ becomes very close to the linear function
$N_{2D}$ in an extended interval of $N_{2D}$. For example, at
$T=4~K$, $N_{spin}(N_{2D})\approx N_{2D}$ for
$0.2*10^{12}<N_{2D}<3*10^{12}$. In the above interval of $N_{2D}$
at $T=4~K$, the hole subsystem is almost completely spin
polarized.

It is interesting to estimate different types of energies
contributing to the formation of stripes. It is easy to see that
the Coulomb energy in eqs.~\ref{TF} dominates the ferromagnetic
and single-particle (kinetic) energies. The Coulomb energy
$U_{Coul}=4\pi e^2d N_{2D}/\epsilon_s\sim 90~meV$ for
$N_{2D}=2*10^{11}~cm^{-2}$ while the spin energy
$|U_{spin}^0|\sim2~meV$ for $N_{spin}=2*10^{11}~cm^{-2}$ at
$T=4~K$. The single-particle kinetic energy under the similar
conditions $E_{kin}\sim n_{2D}/D_{2D}\sim 1.2~meV$. For QDs with
more carriers and higher $n_{2D}$, the above energies become
increased but the condition $U_{Coul}>>|U_{spin}^0|\sim E_{kin}$
remains.

We can also obtain analytic solutions of eqs.~\ref{TF} under
certain conditions. If both spin subsystems of holes
($j_z=\pm3/2$) form a degenerate Fermi gas, the last term in
eqs.~\ref{TF} becomes proportional to the Fermi energy, $2
n_{j_z}/D_{2D}$. Then, we can sum up the equations for
$j_z=\pm3/2$. The resulting equation does not contain the spin
energy $U_{spin}^0$ and can be solved analytically:

\begin{eqnarray}
\label{Na} n_{2D}(r)=\frac{1}{4\pi
e^2d/\epsilon_s+\pi\hbar^2/m_{hh}}[|U_0|-\frac{m_{hh}\omega^2}{2}r^2].
\end{eqnarray}
This formula can be applied, for example, to large QDs in the
spatial region of the ferromagnetic phase F1, $0<r<R_{F1}$, where
$R_{F1}$ is the radius of the F1 phase (see fig.~\ref{fig4}). For
this phase, spin densities can also be found analytically, by
using the condition
$n_{spin}=n_{-3/2}-n_{+3/2}=-N_{spin,saturation}$, where
$N_{spin,saturation}$ is a positive constant equal to $N_{spin}$
at high $N_{2D}$ in fig.~\ref{fig1}b; for $T=4~K$,
$N_{spin,saturation}=3.26~10^{12}~cm^{-2}$ (see fig.~\ref{fig1}b).
The formula (\ref{Na}) also describes the density distribution in
the paramagnetic phase in the regions where the hole gas is
degenerate ($E_f=n_{2D}/D_{2D}>k_BT$). For many other regimes, the
spin densities should be found numerically. Since the Coulomb
energy dominates the magnetic and kinetic terms, the total radius
of a QD can be well estimated from eq.~\ref{Na} by putting
$n_{2D}(R_{QD})=0$. The resulting estimate
$R_{QD}\approx\sqrt{2|U_0|/[m_{hh}\omega^2]}$ is valid at low
temperatures.

Experimentally, the stripe structure of a QD can be observed, for
example, by spatially-resolved optical spectroscopy
\cite{Robinson}. In optical spectroscopy, a spatial resolution can
be as small as $0.1~\mu m$ \cite{Robinson}. Simultaneously optical
emission is sensitive to the spin-polarization of carriers. In an
optical experiment, a ferromagnetic QD would be excited with weak
nonpolarized illumination; the resulting local photoluminescence
will be circularly polarized and reveal the formation of stripes
with different magnetic structures.

Optical methods can also be used to write a magnetic state of QD
(bit: "up"-"down"). This may be done with circularly-polarized
light. Polarized optical pumping can bring a QD into a required
collective magnetic state with spins "up" or "down". In order to
prepare a quantum dot in a required magnetic state, one can also
use a magnetic field induced by an electric current driven through
a metallic wire on the surface of a sample.

Another method to observe the magnetic phases in nano-structures
is the electrical-capacitance spectroscopy
\cite{GovorovCapExper,GovorovCapTheory} which was successfully
applied to observe, for example, compressible and incompressible
stripes in electron quantum wires in the regime of the quantum
Hall effect \cite{GovorovCapExper,GovorovCapTheory}. The
capacitance spectroscopy has been also applied to lateral and
self-assembled quantum dots (see e.g. \cite{QDVoltage2}). For the
nano-structures with relatively large sizes considered in this
paper, the signature of the ferromagnetic phase in the capacitance
spectra is expected to be relatively weak because of the
inequality $U_{Coul}>>|U_{spin}^0|$. However, the formation of the
ferromagnetic phase can be recognized from a critical behavior of
the capacitance spectrum as a function of temperature and voltage.

\section{ Capacitance spectroscopy and magnetic phases in quantum wires}

\begin{figure}[tbp]
\includegraphics*[width=0.6\linewidth,angle=90]{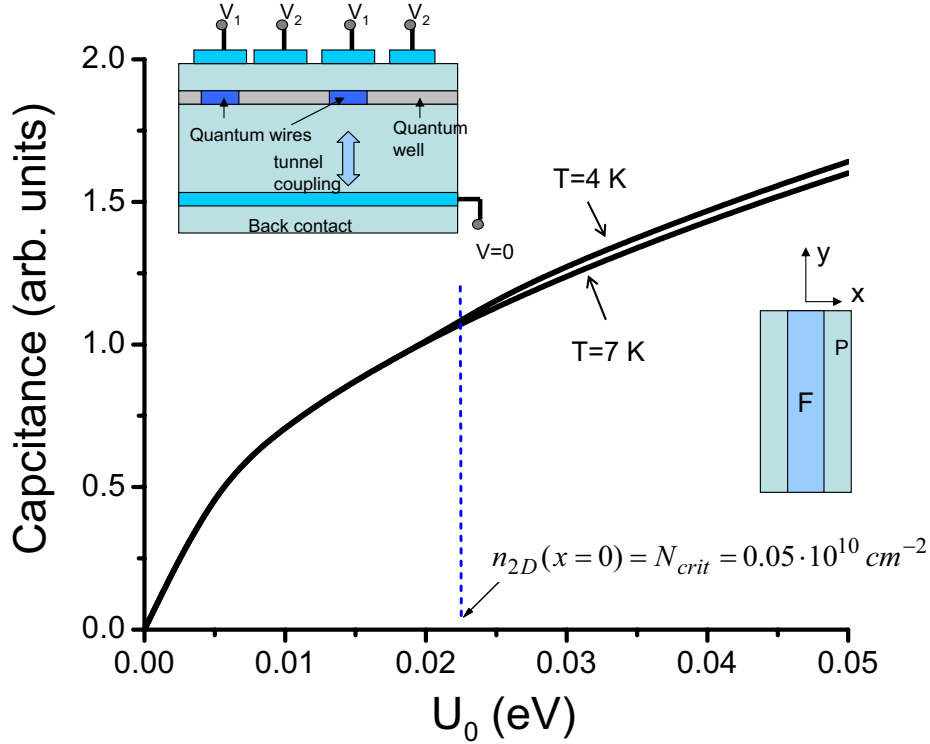}
\caption{Calculated capacitance of a nanowire in the system with
interdigitated metal gate (see inset on the left hand side) for
two temperatures, $4$ and $7~K$. The  inset on the right hand side
shows a top view of a nanowire with two phases.} \label{fig5}
\end{figure}

As an example, we consider here quantum wires in a structure with
the interdigitated metal gate (see inset in fig.~\ref{fig5}). In
such a system, alternating voltages are applied to the metal
strips located on the surface of a quantum well. Similar
structures were studied experimentally in
ref.\cite{GovorovCapExper}. We can calculate the capacitance of a
wire as a derivative $C(V_1)=e (dN_{tot,QW}/dV_1)l\approx
e^2(dN_{tot,QW}/dU_0)l$, where $N_{tot,QW}$ is the linear density
of carriers in a quantum wire and $l$ is the length of a wire in
the in-plane $y$-direction. In the local approximation for the
Coulomb potential (eq.~\ref{FC}), the problems of quantum dot and
wire become similar. Fig.~\ref{fig5} shows the capacitance of a
quantum wire as a function of voltage for two temperatures, just
below and above the Curie temperature $T_C=6.4~K$. One can see in
fig.~\ref{fig5} that at low temperatures ($T<T_C$) the capacitance
becomes increased starting from a critical voltage
$U_0/e\sim0.02~V$. This voltage corresponds to the minimum 2D
density ($\sim5*10^{10}~cm^{-2}$ at $T=4~K$; see fig.~\ref{fig1}b)
which is necessary to obtain the ferromagnetic phase stripe in the
center of nanowire. Starting from this voltage, the central region
of a nanowire contains a ferromagnetic stripe. The capacitance of
a partially ferromagnetic wire becomes increased since the spin
interaction makes a lateral potential well a little deeper and a
wire can accommodate more carriers at a given voltage. If
temperature increases just by $3~K$, the ferromagnetic stripe
vanishes and capacitance becomes reduced. This peculiar
temperature behavior for $U_0>0.02~eV$ can be taken as an evidence
for the ferromagnetic phase.

\section{ Conclusions}

In this paper, we studied quantum dots and wires with many
interacting carriers within the quasi-classical approach. The
strongest interaction in quantum dots with a relatively weak
confinement and a large number of carriers comes from the Coulomb
forces. However, a weaker ferromagnetic interaction determines the
spin structure of a large quantum dot. Depending on the
parameters, a quantum dot can be split into three phases.

\section{ Acknowledgements }

The author would like to thank Bruce McCombe for motivating
discussions. This work was supported by Ohio University and the
A.v. Humboldt Foundation.

\end{document}